\documentclass[twocolumn,nofootinbib,aps,prd,floats,floatfix,amsmath,amssymb,longbibliography,secnumarab
ic,superscriptaddress,preprintnumbers]{revtex4-1} %
\usepackage[final]{graphicx}
\usepackage{amsmath}
\usepackage[usenames]{color}
\usepackage{setspace}
\usepackage{amsmath}
\usepackage{amsthm}
\usepackage{mathrsfs}
\usepackage{epsf}
\usepackage{epsfig}
\usepackage{bbm}
\usepackage{bm}
\usepackage{amsfonts}
\usepackage{amssymb}
\usepackage{latexsym}
\usepackage{graphicx}
\usepackage[english]{babel}
\usepackage{multirow}
\usepackage{float}
\usepackage{url}
\usepackage{slashed}
\usepackage{xcolor} 
\usepackage[utf8]{inputenc}
\usepackage{stmaryrd} 
\usepackage{enumitem}
\usepackage{siunitx}
\usepackage{verbatim}
\usepackage[colorlinks,linkcolor=blue,anchorcolor=blue,citecolor=blue,urlcolor=blue,]{hyperref}
\begin{document}

\newcommand{\eff}{{\text{eff}}}
\newcommand{\SM}{{\text{SM}}}
\newcommand{\BSM}{{\text{BSM}}}
\renewcommand{\topfraction}{0.8}
\newcommand{\fp}[1]{{\bf\color{magenta} fp: #1}}
\newcommand{\tc}{\textcolor{blue}}

\title{Anisotropy of phase transition gravitational wave and its implication for primordial seeds of the Universe}

\author{Yongping Li}
\email{liyp@ihep.ac.cn}
\affiliation{Theoretical Physics Division, Institute of High Energy Physics, Chinese Academy of Sciences,
	19B Yuquan Road, Shijingshan District, Beijing 100049, China}
\affiliation{School of Physics, University of Chinese Academy of Sciences, Beijing 100049, China}

\author{Fa Peng Huang}
\email{Corresponding Author.  huangfp8@sysu.edu.cn}
\affiliation{MOE Key Laboratory of TianQin Mission, TianQin Research Center for Gravitational Physics $\&$ School of Physics and Astronomy, Frontiers Science Center for TianQin, Gravitational Wave Research Center of
CNSA, Sun Yat-sen University (Zhuhai Campus), Zhuhai 519082, China}

\author{Xiao Wang}
\email{wangxiao7@sysu.edu.cn}
\affiliation{MOE Key Laboratory of TianQin Mission, TianQin Research Center for Gravitational Physics $\&$ School of Physics and Astronomy, Frontiers Science Center for TianQin, Gravitational Wave Research Center of
	CNSA, Sun Yat-sen University (Zhuhai Campus), Zhuhai 519082, China}

\author{Xinmin Zhang}
\affiliation{Theoretical Physics Division, Institute of High Energy Physics, Chinese Academy of Sciences,
	19B Yuquan Road, Shijingshan District, Beijing 100049, China}
\affiliation{School of Physics, University of Chinese Academy of Sciences, Beijing 100049, China}

\begin{abstract}
	
We quantitatively study how the primordial density fluctuations	are imprinted on the anisotropy of the phase transition gravitational wave (PTGW).
Generated long before recombination and free from Silk damping, the anisotropic PTGW might reveal the density perturbation seeded from inflation or alternatives. 
We find new behaviors of the PTGW anisotropy power spectrum.
The PTGW anisotropy is stronger than the anisotropy of the cosmic microwave background temperature at all scales, and the high-$\ell$ multiples are enhanced about 1 order due to the early integrated Sachs-Wolfe effect.
Furthermore, differences in primordial power spectra at small scales manifest themselves more significantly on the angular power spectrum of PTGW anisotropy compared to that of the cosmic microwave background.
These properties might provide a novel clue to understanding the primordial density perturbation of our early Universe and thereby complete our understanding of inflation theory.
Taking nanohertz PTGW from dark matter models as a typical example, we obtain amplitudes of PTGW anisotropy which are about 4 or 3 orders weaker than the isotropic PTGW energy spectra.

\end{abstract}

\maketitle

\textit{\textbf{Introduction.}}---
Unraveling the nature of the primordial seeds of our Universe is one of the most important scientific goals in fundamental physics.
The density perturbation from inflation or other alternative theories provides the primordial seeds for the evolution and structure formation of our Universe.
From the anisotropy of the cosmic microwave background (CMB), we could precisely probe the primordial density perturbation at scales $k<0.2~\mathrm{Mpc}^{-1}$.
The Planck satellite has put strong constraints on the power spectrum $C_\ell^{TT}$ of CMB temperature anisotropy in $\ell$ range up to 2500~\cite{Planck:2018vyg}.
However, owing to Silk damping, there is an obvious suppression of the power spectrum at $\ell>1000$. 
Also, the Sunyaev-Zel’dovich effect, cosmic infrared background, and radio source dominate at $\ell > 3000$~\cite{ACTPol:2014pbf}.
New approaches are required to advance our understanding of small-scale density perturbation and the underlying inflation model.

The discovery of gravitational waves (GWs) at LIGO~\cite{LIGOScientific:2016aoc} opens a new window to test the fundamental physics and the problem above.
Among different GW sources, phase transition gravitational waves (PTGWs) \cite{Witten:1984rs,Hogan:1986qda,Kamionkowski:1993fg,Hindmarsh:2013xza} were produced long before CMB photons were last scattered.
Because of its deep connection to dark matter and baryogenesis of the early Universe,
PTGW has attracted a lot of attention and might be tested in various GWexperiments, including LIGO, TianQin~\cite{TianQin:2015yph,TianQin:2020hid}, LISA~\cite{LISA:2017pwj}, Taiji~\cite{Hu:2017mde}, Square Kilometre Array (SKA)~\cite{SKA1:Design}, and FAST~\cite{Nan:2011um}.
Recent studies from NanoGrav~\cite{NANOGrav:2021flc}, PPTA~\cite{Xue:2021gyq}, and EPTA~\cite{Chen:2021rqp} also address possible GW hints that might be explained by PTGW from a dark QCD phase transition.
As in the CMB case, since the isotropic PTGW has been extensively studied, it has become a new and important direction to follow for investigating the anisotropy of PTGWs.
The anisotropic behavior of PTGWs might provide a novel approach to explore the primordial density perturbation, the dynamics of cosmic evolution, the production of dark matter, the origin of baryon-antibaryon asymmetry, the dynamics of the electroweak phase transition, the dark QCD phase transition, the formation of primordial black holes (PBHs), etc.
With the renaissance of PBH, exploring small-scale information becomes attractive since large density perturbation at small scales is needed to form PBH in most models~\cite{Carr:2020xqk}.

References~\cite{Domcke:2020xmn, Jinno:2021ury} study the effects of small-scale density perturbation on the isotropic PTGW energy spectra.
Here we quantitatively investigate the anisotropy of PTGWs.
Reference~\cite{Geller:2018mwu} estimates the anisotropy in the GW background from cosmological phase transition,
Ref.~\cite{Kumar:2021ffi} calculates the primordial non-Gaussianity in the GW background, and Ref.~\cite{Liu:2020mru} studies the large-scale anisotropy of the stochastic GW from domain walls.
In this work, we take the Sachs-Wolfe (SW) and integrated Sachs-Wolfe (ISW) effects into account to study how the primordial density perturbation is imprinted on anisotropic PTGW analytically and numerically for the first time.
Compared to the CMB temperature anisotropy, we find a stronger power spectrum at all scales in our case.
Moreover, in contrast to the damping behavior of CMB, we point out that the power spectrum of PTGW anisotropy gets enhanced by 1 order of magnitude in the high-$\ell$ range because of the ISW effect.
Therefore, the anisotropic PTGW may serve as a new messenger of the primordial density perturbation, especially at small scales.

\textit{\textbf{PTGW from QCD-like phase transition.}}---
We consider the PTGW from a QCD-like phase transition motivated by various dark matter models~\cite{Schwaller:2015tja,Witten:1984rs}.
The first-order phase transition occurring from 1 to 100 MeV can produce PTGWs with peak frequency in the vicinity of nanohertz, which is within the sensitive frequency ranges of pulsar timing array (PTA)~\cite{Hellings:1983fr} experiments such as SKA.
Three mechanisms contribute to PTGWs, namely, bubble collisions~\cite{Huber:2008hg,Kamionkowski:1993fg, Zhong:2021hgo}, sound waves~\cite{Hindmarsh:2013xza,Hindmarsh:2015qta,Hindmarsh:2017gnf}, and magnetohydrodynamic turbulence~\cite{Caprini:2009yp,Kamionkowski:1993fg}.
Recent studies~\cite{Cutting:2018tjt,Cutting:2020nla,Konstandin:2017sat,Lewicki:2020jiv,Lewicki:2020azd, Ellis:2020nnr,Gould:2021dpm,Ellis:2019oqb} give further developments for bubble collision signals.
In most cases, sound wave mechanism is the dominant source of PTGWs~\cite{Hindmarsh:2013xza,Hindmarsh:2015qta,Hindmarsh:2017gnf}. 
Thus, we take PTGWs from the sound wave as an example for simplicity.
The isotropic PTGW energy spectrum from sound wave mechanism with the suppression effect is given by~\cite{Hindmarsh:2017gnf,Wang:2020jrd}
\begin{equation}\label{PTGW:isotropy}
	\begin{aligned}
		h^{2} \Omega_{\mathrm{GW}}(f) \simeq 
		&1.64 \times 10^{-6}\left(\frac{4}{3}\right)^\frac{1}{2}\left(H_{*} R_{*}\right)^{2}\left(\frac{\kappa_{v} \alpha}{1+\alpha}\right)^\frac{3}{2}\\
		&\times\left(\frac{100}{g_{*}}\right)^\frac{1}{3}\left(f / f_{\mathrm{sw}}\right)^{3}\left(\frac{7}{4+3\left(f / f_{\mathrm{sw}}\right)^{2}}\right)^\frac{7}{2}\,\,,
	\end{aligned}
\end{equation}
with the peak frequency
\begin{equation}
	\begin{aligned}
		f_{\mathrm{sw}} \simeq 
		&2.6 \times 10^{-5} \mathrm{~Hz} \frac{1}{H_{*} R_{*}}\left(\frac{T_{*}}{100 ~\mathrm{GeV}}\right)\left(\frac{g_{*}}{100}\right)^\frac{1}{6}\,\,,
	\end{aligned}
\end{equation}
where $\alpha$ is the strength parameter of the phase transition and $g_{*}$ is the total degrees of freedom.
$H_*R_*$ describes the mean bubble separation scaled by the Hubble rate at temperature $T_*$ that the GW generated.
The efficiency parameter $\kappa_{v}$ is the fraction of the energy released by the phase transition that converted into fluid bulk kinetic energy of the plasma and is determined by the bubble wall velocity $v_b$ and $\alpha$~\cite{Espinosa:2010hh}.

The energy spectrum is proportional to the energy density since $H_*^{2}=\rho/3M_{\mathrm{pl}}^{2}$.
We assume here that the density perturbation would produce the anisotropy of the PTGW.
We choose the following benchmark parameters:
\begin{equation}
	\begin{split}
		&\text{Benchmark 1  :  } T_*=1 ~\mathrm{MeV}, ~H_* R_*=0.15\,\,, \\
		&\text{Benchmark 2 :  } T_*=5 ~\mathrm{MeV}, ~H_* R_*=0.2\,\,, \\
	\end{split}
\end{equation}
with $\alpha=0.5$, $v_b= 0.95$, $g_*=10$, and $\kappa_{v}\approx 0.44$.

\textit{\textbf{Anisotropy of the PTGW.}}---
After getting the isotropic PTGW, we calculate its anisotropy.
We employ a line-of-sight integration to describe the free stream of gravitons and calculate the power spectrum of the PTGW anisotropy. 
References~\cite{Contaldi:2016koz,Bartolo:2019oiq,ValbusaDallArmi:2020ifo,Braglia:2021fxn,Ricciardone:2021kel} develop similar methods to calculate the power spectrum of a stochastic GW background.
We expand the distribution function of GWs as
\begin{equation}\label{PTGW:distribution}
	f(\eta,\boldsymbol{x},\boldsymbol{p}) = \bar{f}(\eta,p) - p\frac{\partial\bar{f}(\eta,p)}{\partial p}\mathcal{G}(\eta,\boldsymbol{x},\hat{p})\,\,.
\end{equation}
The dimensionless quantity $\mathcal{G}(\eta,\boldsymbol{x},\hat{p})$ characterizes the perturbation of the distribution function $f$ and gives rise to the anisotropy of PTGWs.
$\boldsymbol{p}$ is the momentum vector of GW. 
Although PTGWs were produced after inflation, the anisotropy is sourced from primordial density perturbation \cite{Geller:2018mwu}. 
Also, the anisotropy we consider here comes from superhorizon scale perturbations at the time of phase transition.
Thus, we assume a frequency independent anisotropy in Eq.~\eqref{PTGW:distribution}.
We use the conformal-Newtonian gauge
\begin{equation}
	ds^2 = -(1+2\Psi)dt^2+a^2(1-2\Phi)\delta_{ij}dx^idx^j\,\,.
\end{equation}
With this metric, the Boltzmann equation then gives the equation that rules the free stream of GWs as
\begin{equation}
	\mathcal{G}^\prime + ik\mu\mathcal{G} =  \Phi^\prime-ik\mu\Psi\,\,.
\end{equation}
A prime denotes the derivative with respect to the conformal time $\eta$. Here $\mu = \hat{k}\cdot\hat{p}$ is the cosine of the angle between $\boldsymbol{p}$ and the Fourier mode $\boldsymbol{k}$.
Integrating over $\eta$ to $\eta_0$, we get the current anisotropy,
\begin{equation}
\begin{aligned}\label{PTGW:integration}
	&\mathcal{G}(\eta_0,k,\mu) \\
	&= \mathcal{G}(\eta_\mathrm{pt},k,\mu) e^{ik\mu(\eta_\mathrm{pt}-\eta_0)}\\
	&~~~ +\int_{\eta_\mathrm{pt}}^{\eta_0} d\eta \left[\Phi^\prime(\eta,k) - ik\mu\Psi(\eta,k)\right] e^{ik\mu(\eta-\eta_0)} \\
	&= \underbrace{\left[\mathcal{G}(\eta_\mathrm{pt},k)+\Psi(\eta_\mathrm{pt},k)\right] e^{ik\mu(\eta_\mathrm{pt}-\eta_0)}}_{\rm SW} \\
	&~~~ +\underbrace{\int_{\eta_\mathrm{pt}}^{\eta_0} d\eta \left[\Phi^\prime(\eta,k) + \Psi^\prime(\eta,k)\right] e^{ik\mu(\eta-\eta_0)}} _{\rm ISW} \,\,.
\end{aligned}
\end{equation}
where $\eta_\mathrm{pt}$ is the conformal time when PTGWs were produced. 
For simplicity, $\mathcal{G}(\eta_\mathrm{pt})$ has no dependence on the GW direction in our assumption.
We integrate the $ik\mu\Psi$ term by parts and drop the $\mu$ independent term $\Psi(\eta_0, k)$, which is an undetectable alteration to the monopole of PTGWs. 
The same procedure can be found in the CMB case, but a surface term is removed because the plasma is extremely opaque to photons, which is not the case for PTGWs.
The initial term for the CMB vanishes for the same reason.
However, in our case, the initial term $\left[\mathcal{G}(\eta_\mathrm{pt},k)+\Psi(\eta_\mathrm{pt},k)\right] e^{ik\mu(\eta_\mathrm{pt}-\eta_0)}$ is what we need and is given by the GW relic from the phase transition at the early stage.
This term shows the SW effect for PTGWs, and the second term gives rise to the ISW effect which comes from the time evolution of the gravitational potential.
These two effects dominate the anisotropy of PTGWs.
As with the last scattering surface of CMB, here we have an emitting surface of PTGWs at time $\eta_\mathrm{pt}\ll \eta_*$.
Free from scattering, PTGWs could carry information from the very early stage closely after inflation.

Now we calculate the anisotropy of $\Omega_{\mathrm{GW}}$ from fluctuation of the distribution function.
The GW energy density can be expressed as
\begin{equation}
\rho_{\mathrm{GW}}(\eta,\boldsymbol{x}) = \int d^3\boldsymbol{p} p f(\eta,\boldsymbol{x},\boldsymbol{p}) = \int dp d\hat{p} p^3 f(\eta,\boldsymbol{x},p,\hat{p})\,\,.
\end{equation}
We also have
\begin{equation}
\rho_{\mathrm{GW}}(\eta,\boldsymbol{x}) = \rho_{c}\int d\ln p \Omega_{\mathrm{GW}}(\eta,\boldsymbol{x},p)\,\,,
\end{equation}
where $\rho_c$ is the critical density of the Universe. 
Thus,
\begin{equation}
\Omega_{\mathrm{GW}}(\eta,\boldsymbol{x},p) = \int d\hat{p}\frac{p^4}{\rho_c}f(\eta,\boldsymbol{x},p,\hat{p})\,\,.
\end{equation}
Then we separate the GW energy spectrum into an isotropic part plus the fluctuation,
\begin{equation}
\begin{split}
\Omega_{\mathrm{GW}}(\eta,\boldsymbol{x},p) 
&= \int \frac{d\hat{p}}{4\pi}\bar{\Omega}_{\mathrm{GW}}(\eta,p)[1+\delta_{\mathrm{GW}}(\eta,\boldsymbol{x},p,\hat{p})]\,\,,
\end{split}
\end{equation}
with the energy density contrast of GW given as
\begin{equation}
	\begin{split}
		\delta_{\mathrm{GW}}(\eta,\boldsymbol{x},p,\hat{p})  
		&=\bigg[4-\frac{\partial \ln \bar{\Omega}_{\mathrm{GW}}(\eta,p)}{\partial \ln p} \bigg] \mathcal{G}(\eta,\boldsymbol{x},\hat{p})\,\,.
	\end{split}
\end{equation}
We define the frequency dependence in the square brackets as $g(p)$ and write in Fourier space
\begin{equation}\label{GWB:fluctuation}
\delta_{\mathrm{GW}}(\eta,k,p,\mu)  
= g(p) \mathcal{G}(\eta,k,\mu)
\end{equation}
Then the power spectrum of GW energy spectrum anisotropy is written as
\begin{equation}\label{freq_dependence}
C_\ell^{\delta_{\mathrm{GW}}}(p) = g^2(p) C_\ell^\mathcal{G}\,\,.
\end{equation}
For PTGWs from sound wave mechanism in the radiation era, $\Omega_{\mathrm{GW}}$ is proportional to the energy density of the radiation.
Thus,
\begin{equation}\label{PTGW:initial_condition}
(\mathcal{G}+\Psi)(\eta_\mathrm{pt},k) = -\frac{1}{3}\mathcal{R}(k)\,\,,
\end{equation}
where $\mathcal{R}$ is the primordial curvature perturbation and is conserved in the superhorizon scale.
All scales we are interested in today were outside the horizon at such an early time.
The SW part of $C_\ell^\mathcal{G}$ can be integrated simply by using the spherical Bessel function $j_\ell$ as follows:
\begin{equation}
\begin{aligned}
C_\ell^{\mathcal{G},\mathrm{SW}} = \frac{2}{9 \pi} \int_{0}^{\infty} d k k^{2} P_{\mathcal{R}}(k)  j_\ell^2[k(\eta_{0}-\eta_\mathrm{pt})]\,\,.
\end{aligned}
\end{equation}
In the following, we use the conventional power-law parametrization $P_{\mathcal{R}}(k)$ at a pivot scale $k_p$  as the default primordial spectrum:
\begin{equation}\label{pps}
P_{\mathcal{R}}(k) = \frac{2\pi^2}{k^3}A_s\left(\frac{k}{k_p}\right)^{n_s-1}\,\,.
\end{equation}
Then,  we have
\begin{equation}
\begin{split}
C_\ell^{\mathcal{G},\mathrm{SW}} =
&2^{n_{s}-2} \frac{\pi^{2}}{9}  A_{s}[ k_{\mathrm{p}}(\eta_{0}-\eta_\mathrm{pt})]^{1-n_{s}}  \\
&\times \frac{\Gamma\left(l+\frac{n_{s}}{2}-\frac{1}{2}\right)}{\Gamma\left(l+\frac{5}{2}-\frac{n_{s}}{2}\right)} \frac{\Gamma\left(3-n_{s}\right)}{\Gamma^{2}\left(2-\frac{n_{s}}{2}\right)}\,\,.
\end{split}
\end{equation}
In a scale-invariant case, $\ell(\ell+1)C_\ell^{\mathcal{G},\mathrm{SW}}/2\pi$ is a constant.

The ISW part of $C_\ell^\mathcal{G}$ has no simple analytic expression, and we estimate it at large $\ell$ as follows. 
The multiple moment is
\begin{equation}
\mathcal{G}_\ell^{\mathrm{ISW}}(\eta_0,k) = \int_{\eta_{p t}}^{\eta_{0}} d \eta\left( \Phi^{\prime}+\Psi^{\prime} \right) \left(\eta, k\right) j_{\ell}\left[k\left(\eta_0-\eta\right)\right]\,\,.
\end{equation}
A perturbation with wave number $k$ contributes most strongly to multiples $\ell\sim k(\eta_0-\eta_\mathrm{pt})$. 
For a large-scale $k(\eta_0-\eta_\mathrm{pt})\ll\ell$, the peak of $j_\ell$ is always far outside the integration range. 
Thus, the transfer function is approximately zero.
We only consider wave number range $k(\eta_0-\eta_\mathrm{pt})\gtrsim\ell$. 
For small-scale $k(\eta_0-\eta_\mathrm{pt}) \gtrsim \ell\gg 1$ perturbations decay rapidly after entering the horizon at a very early stage.
Thus, we can approximate as
\begin{equation}
\left(\Phi^\prime+\Psi^\prime\right)(\eta,k) \approx -(\Phi+\Psi)(\eta_\mathrm{pt},k)\delta(\eta-\eta_k)\,\,,
\end{equation}
where $\eta_k$ is the conformal time when the $k$ mode decays.
Phase transition happened long before matter-radiation equality, so we set $\eta_\mathrm{pt}$ to zero for our integration. 
The above approximation is valid because the peak width of $j_\ell$ is about $k\Delta\eta\sim l^{1/3}$, which is much larger than the peak width of $\left(\Phi^\prime+\Psi^\prime\right)$. 
Thus, we get the following final expression for multiples at large $\ell$
\begin{equation}
\mathcal{G}_\ell(\eta_0,k) \approx [(\mathcal{G}+\Psi)-(\Phi+\Psi)]\left(\eta_{p t}, k\right)j_\ell[k(\eta_{0}-\eta_\mathrm{pt})]\,\,.
\end{equation}
Then the initial condition is
\begin{equation}
[(\mathcal{G}+\Psi)-(\Phi+\Psi)]\left(\eta_{p t}, k\right) = \mathcal{R}(k) \quad\text{for large $\ell$}.
\end{equation}
Thus, the overall power spectrum at large $\ell$ is 9 times larger than $C_\ell^{\mathcal{G},\mathrm{SW}}$ only.

With the power spectrum, we can compute the variance of the fluctuation with
\begin{equation}
	\begin{split}
		\mathrm{Var}^\mathcal{G}
		&= \frac{1}{4\pi}\sum_\ell(2\ell+1)C_\ell^{\mathcal{G}}\,\,.
	\end{split}
\end{equation}
For $\delta_{\mathrm{GW}}(\eta_{0},\boldsymbol{x}_0,p,\mu)$, we have $\mathrm{Var}^{\delta_\mathrm{GW}}(p)= {g^2(p)}\mathrm{Var}^\mathcal{G}$.
We employ the frequency dependent standard deviation
\begin{equation}
 \sigma_{\mathrm{GW}}(p) \equiv h^2\Omega_{\mathrm{GW}}(p)\sqrt{\mathrm{Var}^{\delta_\mathrm{GW}}(p)}
\end{equation}
to quantify the amplitude of the PTGW anisotropy.
It is worth noticing that one should use the normalized variance $\mathrm{Var}^\mathcal{G}$ for an evaluation of the density perturbation.
\begin{figure}[!htbp]
	\begin{center}
		\includegraphics[width=0.45\textwidth,clip]{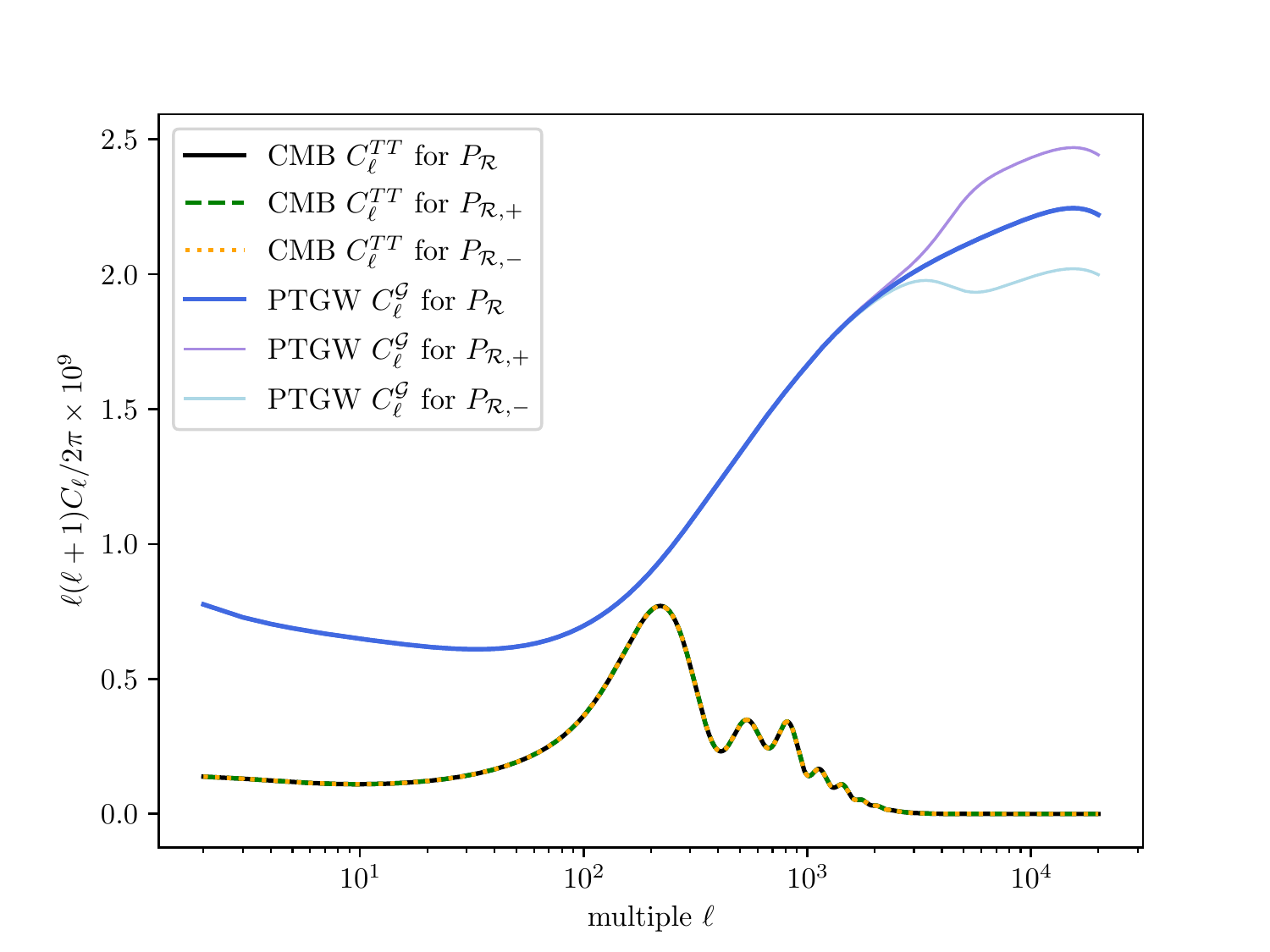}
		\caption{\textbf{Angular Power Spectrum}: The blue line is angular power spectrum $C_\ell^\mathcal{G}$ for the anisotropy of PTGW with the  default primordial power spectrum $P_\mathcal{R}$ given by parameters from Planck~\cite{Planck:2018vyg}. The lightpurple line is for $P_\mathcal{R,+}$ with the amplitude being increased by 10 percent in $k>0.3~\mathrm{Mpc}^{-1}$ range compared to $P_\mathcal{R}$, and the lightblue one is for $P_\mathcal{R,-}$ with amplitude decreased by 10 percent in $k>0.3~\mathrm{Mpc}^{-1}$ range. The black solid line, green dashed line and orange dotted line are CMB temperature power spectra $C_\ell^{TT}$ corresponding to $P_\mathcal{R}$, $P_\mathcal{R,+}$ and $P_\mathcal{R,-}$ separately. We can see obvious distinctions in $C_\ell^\mathcal{G}$ for PTGW, but not in $C_\ell^{TT}$ for CMB. All angular power spectra are dimensionless, so one should multiply $C_\ell^{TT}$ by the squared CMB temperature to get the conventional one in unit of $\mathrm{\mu K}^2$.}
		\label{power_spectrum}
	\end{center}
\end{figure}
\begin{figure}[!htbp]
	\begin{center}
		\includegraphics[width=0.4\textwidth,clip]{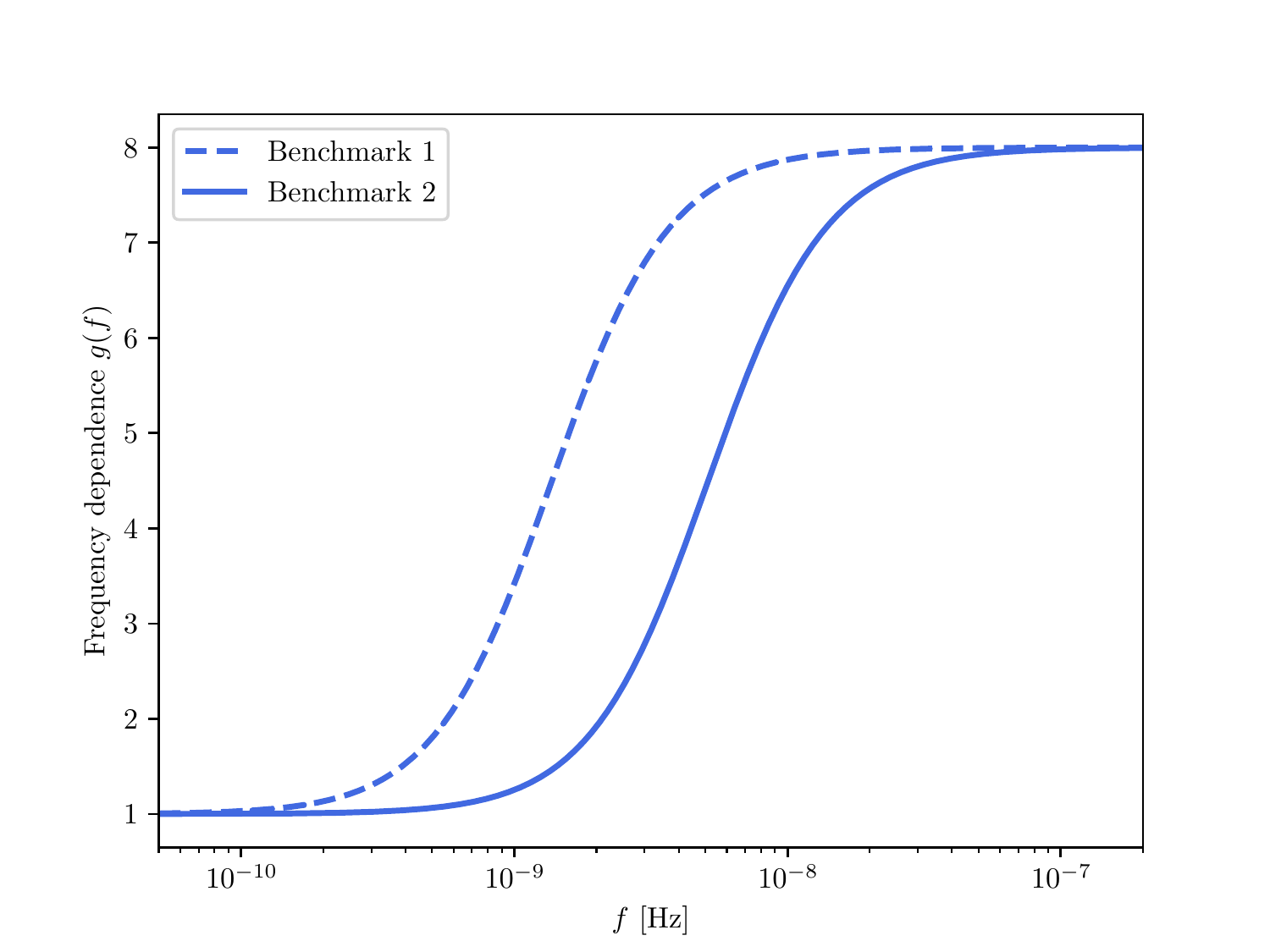}\\
		\includegraphics[width=0.4\textwidth,clip]{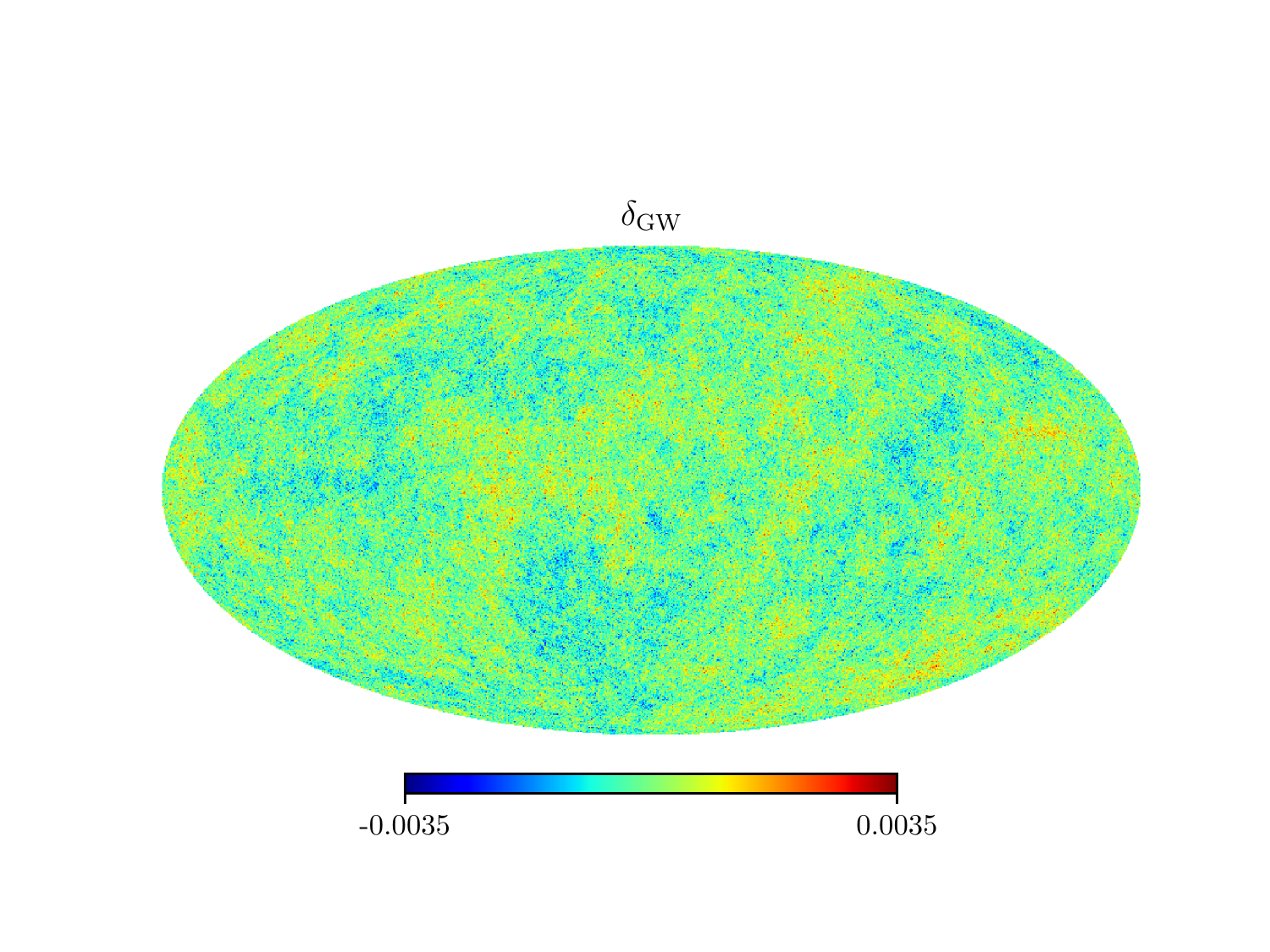}\\
		\caption{\textbf{Upper}: The frequency dependence $g(f)$ between the fluctuation $\delta_{GW}$ of the energy spectrum $\Omega_{\mathrm{GW}}$ and the fluctuation $\mathcal{G}$ of the distribution function. The variable momentum $p$ has been substituted with frequency $f$. The dashed line is $g(f)$ for benchmark 1 and the solid one is for benchmark 2. \textbf{Lower}: Realization of PTGW anisotropy power spectrum $C_\ell^{\delta_{\mathrm{GW}}}$.}
		\label{map}
	\end{center}
\end{figure}
\begin{figure}[!ht]
	\begin{center}
		\includegraphics[width=0.45\textwidth,clip]{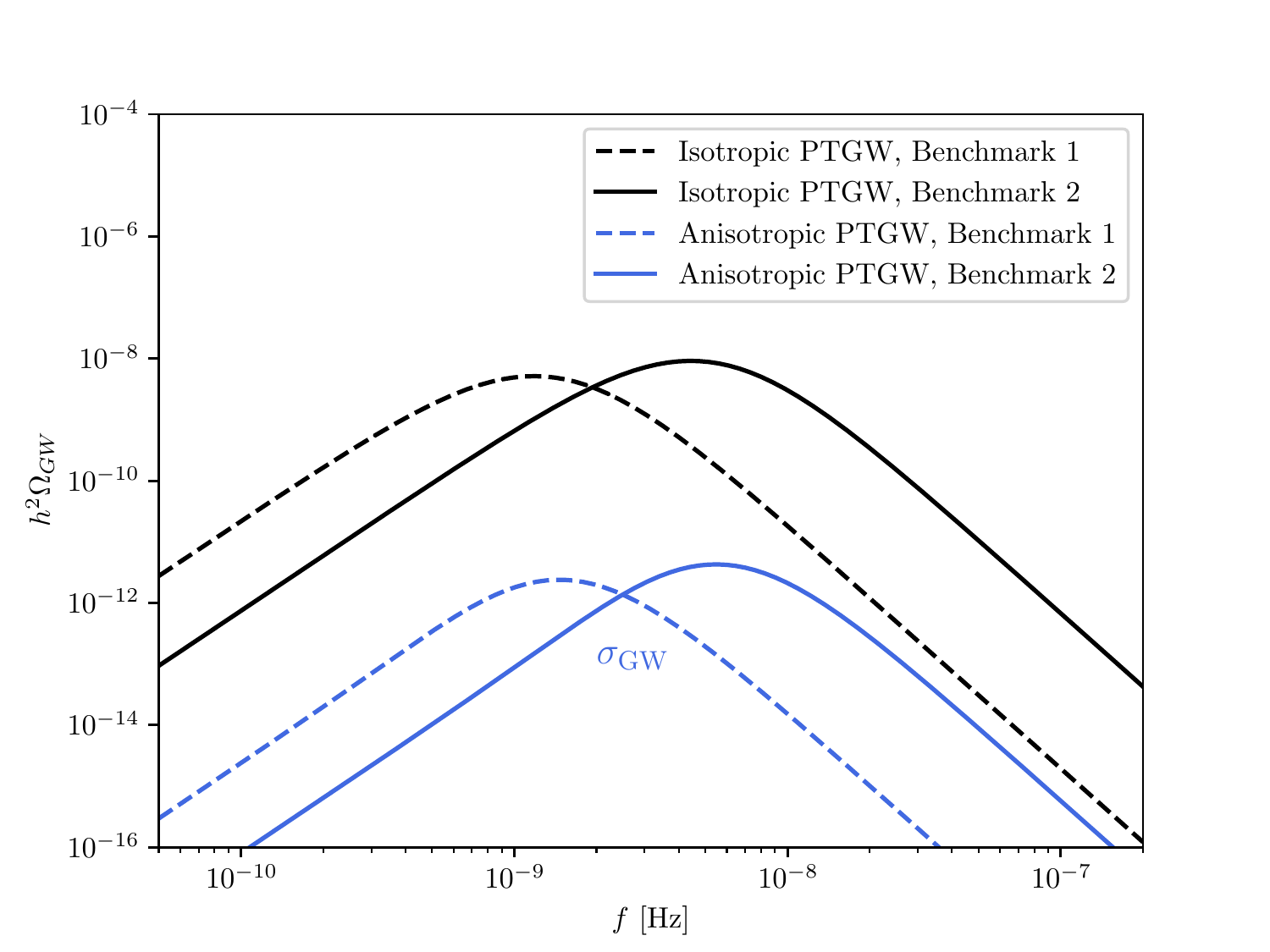}\\
		\caption{\textbf{The isotropic and anisotropic PTGW energy spectra for the
				default primordial power spectrum}: The black lines are the isotropic PTGW, dashed one for benchmark 1 and solid one for benchmark 2. The blue lines are the anisotropic PTGW,  dashed one for benchmark 1 and solid one for benchmark 2. The amplitude of the anisotropy is about four or three orders weaker than the isotropic PTGW.}
		\label{sensitivity}
	\end{center}
\end{figure}

\textit{\textbf{Results and Discussion.}}---
Planck~\cite{Planck:2018vyg} currently gives us the best constraint on the primordial power spectrum, and we take this as the starting point: $\ln(10^{10}A_s)=3.044$ and $n_s=0.966$. 
The pivot scale is set at $k_p=0.05~\mathrm{Mpc}^{-1}$. 
We modify $\mathtt{CLASS}$~\cite{Blas:2011rf} to compute the power spectrum and testify our analytical calculation.
The angular power spectra of PTGW anisotropy and CMB for different primordial power spectra are shown in Fig.~\ref{power_spectrum}.
Note that since all the power spectra that we plot are dimensionless, one should multiply $C_{\ell}^{TT}$ by the square of the background temperature of CMB to get the most commonly used one in the literature.

We show in Fig.~\ref{power_spectrum} the differences between the PTGW and CMB angular power spectra for the default primordial power spectrum $P_\mathcal{R}(k)$ as given in Eq.~(\ref{pps}) with parameters from Planck.
The blue line is $\ell(\ell+1)C_\ell^\mathcal{G}/2\pi$ for PTGW anisotropy, and the black solid line is the CMB temperature spectrum.
We find that, at all scales, the PTGW shows stronger anisotropy than the CMB temperature.
The most obvious feature of $C_\ell^\mathcal{G}$ is that it goes up instead of damping quickly at high $\ell$. 
This is a result of the ISW effect and the absence of Silk damping.
The ISW effect contributes mostly to the rising of angular power spectrum at high $\ell$.
GWs can hardly be scattered by matter and dark matter, meaning there is no Silk damping.
Thus, PTGW anisotropy keeps the small-scale information from the early Universe which was erased in the CMB.

We then slightly vary the primordial power spectrum at small scales and show how the response on the angular power spectrum of the PTGW differs from that of the CMB.
We increase and decrease the primordial power spectrum by $10\%$ starting from a step at $k_s=0.3~\mathrm{Mpc}^{-1}$ by taking the following parametrization:
\begin{equation}
	P_\mathcal{R,\pm}(k)= P_\mathcal{R}(k)\left(1\pm 0.1\times\frac{\tanh[B(k-k_s)]+1}{2}\right)\,\,,
\end{equation}
with $P_\mathcal{R}(k)$ being the default primordial power spectrum and $B=10~\mathrm{Mpc}$.
As shown in Fig.~\ref{power_spectrum}, the light purple line is the PTGW power spectrum for $P_\mathcal{R,+}$ and the light blue line is for $P_\mathcal{R,-}$. 
We can see that these two lines deviate obviously at high $\ell$ from $C_\ell^\mathcal{G}$ for the default power spectrum $P_\mathcal{R}$. 
The green dashed line and the orange dotted line are the CMB temperature power spectrum corresponding to $P_\mathcal{R,+}$ and $P_\mathcal{R,-}$, respectively.
We can see that they almost overlap with the one for $P_\mathcal{R}$.
In summary, remarkable distinctions can be seen in PTGW power spectra, while no significant difference appears in CMB temperature angular power spectra; small-scale information is retained more in PTGW anisotropy.
To pin down an inflation model requires a solid understanding of the small-scale primordial power spectrum; different models might give different small-scale behaviors.
See Ref.~\cite{Chluba:2012we} and references therein.
Thus, PTGWs could theoretically be helpful to distinguish among different inflation models.

The frequency dependence factor $g(f)$ between $\delta_{\mathrm{GW}}$ and $\mathcal{G}$ for the default primordial power spectrum is shown in the upper panel of Fig.~\ref{map}. 
The momentum $p$ is substituted by frequency $f$ hereafter.
The dashed and solid lines are for benchmarks 1 and 2, respectively.
This factor equals 8 in the frequency range that is larger than the peak frequency $f_{\mathrm{sw}}$. 
We absorb it into $C_\ell^{\delta_{\mathrm{GW}}}$ as in Eq.~(\ref{freq_dependence}) and give one realization map as an example in the lower panel of Fig.~\ref{map}. 
We remind the reader that the map here is not the real distribution but instead only a realization of our PTGW power spectrum.
The real one might be correlated to the CMB temperature field because the two share the same origin, which is the primordial density perturbation.

The amplitude of anisotropic PTGW for the default primordial power spectrum is shown in Fig.~\ref{sensitivity}. 
The black lines are the isotropic PTGW for benchmark 1 (dashed) and 2 (solid).
The blue lines are the anisotropic PTGW, dashed for benchmark 1 and solid for benchmark 2.
The blue lines show $\sigma_{\mathrm{GW}}$ at different frequencies. 
The shape of $\sigma_{\mathrm{GW}}$ differs from the isotropic PTGW energy spectrum $h^2\Omega_{\mathrm{GW}}$ because of the factor $g(f)$ shown in the upper panel of Fig.~\ref{map}. 
The shape change is small and not significant in the log-scale plot of the power spectrum.
Compared to the $4\times10^{-5}$ level CMB temperature anisotropy (which can be computed from the CMB temperature map from Planck or the results of the power spectrum), the anisotropic PTGW is at the $1\times10^{-4}$ level for $\mathcal{G}$ and the $8\times10^{-4}$ level for the frequency dependent $\delta_{\mathrm{GW}}(f)$ in the range right to the peak frequency.
The measured variance is related to the resolution of the observation, but the small-$\ell$ multiples contribute the most.
So the value of $\sigma_{\mathrm{GW}}$ does not change significantly if the resolution is not far from the degree level.

Detailed study of the detectability of the anisotropy components is an important but challenging new direction which is beyond the scope of this work.
For PTA observation, the GW affects the pulsar timing residuals, and the relative direction between GW propagation and pulsar location determines the response on timing residuals.
Known locations of pulsars, a specific distribution of GW strength on the sky, leave a unique imprint on the crosscorrelation of timing residuals between pulsars, which is called the overlap reduction function.
Thus, information about PTGW anisotropy as well as isotropy is coded into the overlap reduction function.
The famous Hellings and Downs curve is the overlap reduction function for an isotropically distributed GW background. 
With knowledge about pulsars and an assumption of GW strength angular distribution, a prior of the overlap reduction function can be achieved.
The prior goes into the covariance matrix of the timing residuals of the pulsars, and one then fits the data with the pulsar timing model and covariance matrix; 
see Refs.~\cite{Taylor:2013esa,Mingarelli:2013dsa,Romano:2016dpx,Cornish:2014rva,Gair:2014rwa,Ali-Haimoud:2020ozu,Ali-Haimoud:2020iyz} for details. 
In general, a PTA with a pulsar population of N can constrain at most $N_{\rm pair}=N(N-1)/2$ spherical harmonic amplitudes, which corresponds to $\ell_{\rm max}\le \sqrt{N_{\rm pair}}-1\approx (N-1)/\sqrt{2}-1$. 
Future PTA experiments such as SKA may have the potential to observe $\sim 10^4$ pulsars, which corresponds roughly to $\ell_{\mathrm{max}}\sim 10^4$. 
However, according to current studies on PTA data analysis, the detection will be challenging because the sensitivity for anisotropy is more downgraded at higher multiples.
We are expecting future GW experiments with better sensitivity and resolution, data combined from different experiments \cite{Garcia-Bellido:2021zgu}, and also new methods to explore the anisotropy of PTGWs.

\textit{\textbf{In conclusion}}, 
We have developed a methodology for computing the anisotropy of phase transition gravitational waves from sound wave mechanism.
We calculated the power spectrum of the anisotropy analytically and numerically.
The Sachs-Wolfe effect is considered, and we find that a higher level of anisotropy at all scales shows up than in the CMB case.
The power spectrum was even enhanced by about 1 order of magnitude at high $\ell$ due to the integrated Sachs-Wolfe effect.
Different primordial power spectra at small scales could lead to different PTGW angular power spectra, but it could lead to almost the same CMB angular power spectrum.
Thus, the anisotropy may leave us a window to explore the primordial density perturbation, especially at small scales.
Our results can be applied to phase transitions at different energy scales, which would trigger gravitational waves with different peak frequencies.
We leave detailed discussions on the anisotropy of gravitational waves by bubble collision and turbulence to future work.
We point out that, for these two mechanisms, the integrated Sachs-Wolfe effect also contributes a large portion to small-scale anisotropy.
Besides the implications for dark matter or baryogenesis in the early Universe from the isotropic phase transition gravitational wave, the study of its anisotropy here might be helpful for understanding the underlying inflation or alternative theories.

We thank Hong Li, Siyu Li, Yang Liu, and Zhiqi Huang for the useful discussions. Y. L. and X. Z. are supported by the National Natural Science Foundation of China (Grant No. 11653002) and the National Key R\&D Program of China (Grant No. 2020YFC2201600). F. P. H. and X. W. are funded by the Guangdong Major Project of Basic and Applied Basic Research (Grant No. 2019B030302001).

\bibliographystyle{apsrev4-1}
\bibliography{reference}
\end{document}